\documentclass[11pt]{article}
\usepackage[final]{acl}
\usepackage{times}
\usepackage{latexsym}
\usepackage[T1]{fontenc}
\usepackage[utf8]{inputenc}
\usepackage{microtype}
\usepackage{inconsolata}
\usepackage{graphicx}
\usepackage{enumitem}
\usepackage{placeins}
\usepackage{float}
\usepackage{booktabs}
\title{Adversarial Network Imagination: Causal LLMs and Digital Twins for Proactive Telecom Mitigation}
\author{
  Vignesh Sriram \quad
  Yuqiao Meng \quad
  Luoxi Tang \quad
  Zhaohan Xi \\
  Binghamton University
}

\begin{document}
\maketitle
\begin{abstract}
Telecommunication networks experience complex failures such as fiber cuts, traffic overloads, and cascading outages. Existing monitoring and digital twin systems are largely reactive, detecting failures only after service degradation occurs. We propose \emph{Adversarial Network Imagination}, a closed-loop framework that integrates a Causal Large Language Model (LLM), a Knowledge Graph, and a Digital Twin to proactively generate, simulate, and evaluate adversarial network failures. The Causal LLM produces structured failure scenarios grounded in network dependencies encoded in the Knowledge Graph. These scenarios are executed within a Digital Twin to measure performance degradation and evaluate mitigation strategies. By iteratively refining scenarios based on simulation feedback, the framework shifts network operations from reactive troubleshooting toward anticipatory resilience analysis.
\end{abstract}

\section{Introduction}

Telecommunication networks constitute critical infrastructure, supporting mobile connectivity, cloud services, and latency-sensitive applications. As these networks grow in scale and complexity, failures such as link outages, traffic surges, and misconfigurations can propagate rapidly across interconnected components, leading to widespread service disruption. Although modern networks are extensively monitored, most mitigation mechanisms remain reactive, responding only after performance degradation or service loss occurs.

Digital twin technologies have emerged as a promising paradigm for modeling, monitoring, and optimizing complex networked systems, particularly in 5G and beyond networks \citep{Ahmed2022,Ayoubi2019,Tao2019}. Prior work has leveraged digital twins to simulate network topology, routing behavior, and resource utilization for offline diagnosis and
planning \citep{Ahmed2022,DeOliveira2022,Zhang2021}. However, these approaches typically rely on manually crafted or historically observed
failure scenarios. Similarly, fault-injection frameworks simulate predefined disruptions, but struggle to capture
rare, coordinated, or cascading failures arising from complex dependency structures.

Causal modeling and knowledge graphs have been applied to represent dependencies among network components and improve fault localization
\citep{Pearl2018,Kaur2021}. While effective for diagnosis and explanation, these methods are largely non-generative and do not support
the proactive exploration of counterfactual or adversarial failure scenarios. As a result, existing systems lack a principled mechanism to
systematically imagine and evaluate novel failure patterns before they manifest in real networks.

Recent advances in large language models (LLMs) offer new opportunities for structured reasoning and multi-step generation. However, unconstrained language generation often produces scenarios that violate causal structure or describe infeasible interventions, limiting its applicability to safety-critical domains such as telecommunications. This motivates the question of how LLMs can be guided to generate hypothetical failure scenarios that are both causally consistent and operationally meaningful.

\begin{figure}[t]
\centering
\includegraphics[width=\linewidth]{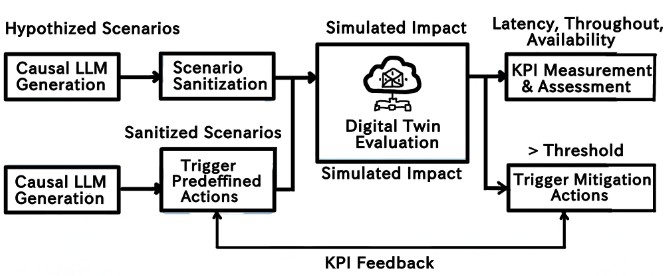}
\caption{Overview of the proposed Adversarial Network Imagination framework.}
\label{fig:intro_overview}
\end{figure}

To address this challenge, we propose \emph{Adversarial Network
Imagination}, a proactive framework that integrates causal large language models, knowledge graphs, and a high-fidelity digital twin. As illustrated in Figure~\ref{fig:intro_overview}, a Causal LLM generates structured and explainable failure scenarios grounded in a dependency graph encoding network topology and resource relationships. These scenarios are executed
within a digital twin to simulate cascading effects and evaluate mitigation strategies, with simulation feedback used to iteratively
refine scenario generation in a closed loop.

By grounding language generation in explicit causal structure and validating outputs through simulation, the proposed approach enables
anticipatory analysis of complex failure dynamics that are difficult to enumerate manually. While we focus on telecom networks, the formulation of causally constrained scenario generation and the evaluation methodology are applicable to other structured, safety-critical systems.

This paper makes the following contributions:
\begin{itemize}
    \item We formulate \textbf{causal scenario generation} as a constrained
    language generation problem in which LLM outputs must respect explicit
    dependency graphs and intervention semantics.
    \item We define a \textbf{Causal LLM} whose generation is grounded in
    causal structure to produce hypothetically valid, multi-step failure
    scenarios.
    \item We propose a closed-loop framework in which a digital twin acts
    as an external verifier, enabling iterative refinement of
    LLM-generated scenarios.
    \item We empirically demonstrate that causal constraints improve
    scenario validity, cascading depth, and downstream mitigation
    effectiveness compared to unconstrained and rule-based baselines.
\end{itemize}

\section{Related Work}

Our work relates to research on causal modeling and knowledge graphs for
representing dependencies in complex systems, as well as recent advances
in large language models for reasoning and structured generation.

\subsection{Causal Modeling and Knowledge Graphs}

Causal modeling provides a principled framework for representing
dependencies and reasoning about interventions and counterfactuals
\citep{Pearl2018}. Recent work has explored causal
representation learning and the integration of causal structure with machine learning models. Knowledge graphs offer a complementary abstraction to encode structured relationships and have been applied to fault diagnosis, dependency analysis, and root-cause localization in large-scale systems and networks
\citep{Kaur2021,Zhang2021}.

In networks and systems domains, causal and graph-based models have
been used to improve fault localization and explainability
\citep{Ayoubi2019,Ahmed2022}. However, these approaches typically rely on
predefined failure patterns or historical observations and lack generative mechanisms to actively explore novel or adversarial
failure scenarios. Our work differs by leveraging explicit causal structure to guide generative scenario synthesis under structured
dependency constraints.

\subsection{LLMs for Reasoning and Structured Generation}

Large language models have demonstrated strong reasoning and multi-step generation capabilities \citep{Wei2022,Chowdhery2022}, with techniques such as chain-of-thought prompting, tool use, and constrained decoding
improving logical consistency and interpretability
\citep{Yao2023,Schick2023}. Recent work has further explored incorporating external knowledge, constraints, or feedback signals to guide generation \citep{Madaan2023,Bai2022}.

While LLMs have been applied to systems modeling and simulation, most
existing approaches operate at the level of unconstrained text generation or sequence-level evaluation and do not explicitly enforce
causal validity with respect to structured system dependencies. In contrast, our approach integrates causal constraints and simulation
feedback directly into the generation process, enabling coherent multi-hop reasoning over complex networked systems.

\section{Background and Problem Definition}

Failure propagation in large-scale networked systems has been widely studied using graph-based and probabilistic models, with prior work
examining cascading outages and dependency-driven disruptions. Related research in network tomography has explored
inferring internal failures from end-to-end measurements, but such approaches often struggle to capture complex multi-hop dependencies \citep{Nguyen2016}.

Telecommunication networks consist of interconnected routers, links, and services whose dependencies can cause localized failures—such as fiber cuts, router overloads, misconfigurations, or attacks—to propagate
rapidly and impact large portions of the network. Traditional testing methods, including rule-based fault injection and replay of historical incidents, are limited in their ability to explore rare, coordinated, or
multi-step failure scenarios, while reactive monitoring detects failures only after service degradation has occurred.

\begin{figure}[h!]
\centering
\includegraphics[width=0.9\linewidth]{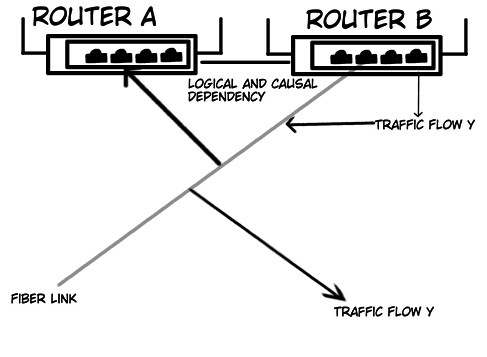} 
\caption{Mini Knowledge Graph of a telecom network illustrating routers, services, traffic flows, and their dependencies.}
\label{fig:knowledge_graph}
\end{figure}

To address these limitations, we combine a structured Knowledge Graph with causal reasoning within a digital twin environment. The Knowledge Graph encodes network topology, routing relationships, service dependencies, and shared-resource constraints, enabling explicit modeling of how failures propagate across components. This representation supports the systematic exploration of hypothetical
failure scenarios and proactive mitigation analysis. Figure~\ref{fig:knowledge_graph}
illustrates an example Knowledge Graph for a representative network topology.

\section{Method}

\begin{figure*}[t]
\centering
\includegraphics[width=0.75\textwidth]{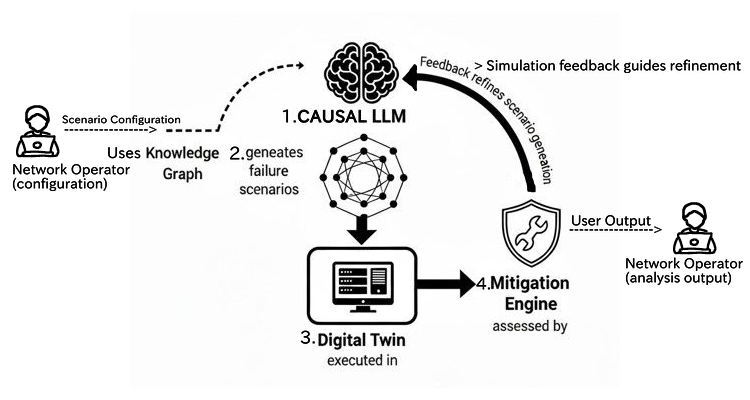}
\caption{Overview of the Adversarial Network Imagination framework. The Causal LLM generates failure scenarios using the Knowledge Graph, which are executed in the Digital Twin and assessed by the Mitigation Engine. Feedback refines scenario generation in a closed loop. Operators interact with the framework by configuring scenario constraints and inspecting simulated outcomes, while all failure generation and evaluation remain fully automated within the causal loop.}
\label{fig:method_overview}
\end{figure*}

This section describes the design and workflow of the proposed \emph{Adversarial Network Imagination} framework, which enables proactive failure analysis and mitigation in telecom networks. The framework integrates causal reasoning, generative modeling, and high-fidelity simulation to systematically explore and evaluate complex network failures. Figure~\ref{fig:method_overview} provides an overview of the full workflow, showing how the components interact to generate, simulate, and assess adversarial scenarios.

\subsection{Causal Language Modeling}
Standard LLMs generate text by modeling surface-level token dependencies, which can lead to logically inconsistent or causally implausible outputs when reasoning about structured systems. In contrast, a Causal LLM
operates under explicit constraints derived from a dependency graph, restricting generation to interventions and effects supported by causal
structure.

In this work, causal structure is provided externally via a knowledge graph encoding component dependencies. During generation, the LLM is prompted with graph context and required to produce event sequences that satisfy the causal consistency conditions defined in Definition~1. This enables counterfactual reasoning: the model generates hypothetical
interventions that may not appear in historical data but remain structurally valid.

Unlike post-hoc filtering, causal constraints are enforced during generation, reducing hallucinated or infeasible scenarios and improving the reliability of downstream reasoning.

\subsection{Causal LLM}
\label{sec:causal_llm}

We propose a \emph{Causal LLM} for generating network failure scenarios under explicit dependency constraints derived from a network knowledge graph. Unlike standard language models that optimize unconstrained sequence-level likelihood, the Causal LLM conditions generation on structured causal information, enabling coherent multi-step reasoning
over network components and their dependencies.

\textbf{Causal Conditioning.}
The model is grounded in a knowledge graph encoding network topology, routing dependencies, shared resources, and service relationships. During generation, the LLM is provided with the relevant subgraph and an
explicit intervention context, allowing it to reason about which components can plausibly affect others rather than relying solely on
surface-level language patterns.

\textbf{Constrained Generation.}
Scenario generation is constrained so that each produced event corresponds to a valid intervention on a network component and remains
causally consistent with preceding events. Infeasible actions—such as affecting unreachable components or violating implied temporal precedence—are disallowed. These constraints ensure that generated
scenarios form executable, multi-step failure chains rather than isolated or implausible descriptions.

\textbf{Counterfactual Reasoning.}
Rather than replaying historical failures, the Causal LLM generates counterfactual scenarios that may not have previously occurred but
remain causally valid. This enables systematic exploration of rare, adversarial, or cascading failure patterns that are difficult to
enumerate manually, while avoiding hallucinated or logically inconsistent outcomes.

\textbf{External Verification Interface.}
Generated scenarios are executed within a digital twin, which serves as an external verifier by simulating system behavior and recording resulting impacts. While simulation feedback is used to refine future
generation, the LLM itself remains responsible for producing causally coherent event sequences. This separation allows language-level reasoning quality to be evaluated independently from system performance.

Overall, the Causal LLM transforms free-form language generation into a structured mechanism for imagining and evaluating hypothetical network failures under explicit causal constraints.

\subsection{Knowledge Graph}
The Knowledge Graph encodes the structural and functional dependencies of the telecom network, including topology links, routing paths, shared resources (e.g., CPU, memory, buffer capacities), and historical incidents. It provides a structured context for the Causal LLM to ensure that generated failures obey real-world constraints, such as component reachability, service dependencies, and load-sharing relationships. The graph also supports causal reasoning, allowing the LLM to infer indirect effects and cascading failure patterns that may emerge from primary disruptions. The causal modeling approach adopted in this work is inspired by foundational principles of causal inference, which emphasize reasoning about interventions and counterfactual outcomes rather than surface-level correlations \citep{Pearl2018}.

\subsection{Adversarial Scenario Generator}
The Adversarial Scenario Generator refines and organizes LLM outputs into executable sequences for simulation. It introduces complexity by combining multiple failure events, adjusting temporal ordering, and injecting realistic stress conditions such as traffic surges or simultaneous device outages. This component ensures that scenarios are feasible for simulation while maintaining the adversarial intent necessary to test network resilience.

\subsection{Digital Twin Execution Engine}
The Digital Twin simulates the network under the proposed failure scenarios, providing a safe and controlled environment for evaluation. It models routing, resource utilization, congestion, and service-level performance, capturing both immediate and cascading effects. Performance metrics—including latency, packet loss, reroute time, and impacted nodes—are recorded for each scenario. This step allows the framework to quantify system vulnerability and validate whether proposed failures lead to meaningful degradation.

\subsection{Mitigation Engine}
The Mitigation Engine evaluates candidate recovery strategies for each simulated scenario. It applies corrective actions such as traffic rerouting, resource reallocation, or rollback procedures, and measures their effectiveness in restoring stable network operation. Feedback from the mitigation process is fed back into the LLM and scenario generator, enabling iterative refinement of scenario generation and enhancing the framework’s ability to uncover complex, high-impact failures.

\subsection{Workflow Summary}
Overall, the method implements a closed-loop workflow in which failure scenarios are generated, simulated, and assessed in a continuous cycle. The integration of causal reasoning, knowledge-guided generation, and high-fidelity simulation allows operators to anticipate and mitigate failures proactively, shifting telecom networks from reactive response toward anticipatory resilience analysis.

\section{Experiments}

Due to the exploratory and architectural nature of this work, our evaluation emphasizes scenario diversity, causal validity, and mitigation effectiveness rather than large-scale quantitative benchmarking. We evaluate whether the proposed framework can generate realistic, causally grounded failure scenarios, simulate propagation effects within a Digital Twin, and support informative mitigation analysis under adversarial conditions.

\subsection{Experimental Settings}

\textbf{Datasets.} 
We use publicly available network topology and traffic datasets commonly adopted in networking research. ISP-level and backbone topologies are sourced from the Internet Topology Zoo, while AS-level dependency structures are informed by CAIDA datasets. Adversarial and bursty traffic patterns are derived from MAWI backbone traces. These datasets are used to instantiate Knowledge Graphs and Digital Twin environments rather than for supervised training.

\begin{table*}[t]
\centering
\small
\resizebox{1.3\columnwidth}{!}{%
\begin{tabular}{lcccccl}
\toprule
\textbf{Dataset} & \textbf{Type} & \textbf{Nodes} & \textbf{Links} & \textbf{Failures} & \textbf{Traffic} & \textbf{Purpose} \\
\midrule
TopologyZoo-1 & ISP Backbone & 120 & 180 & Link, Router & Normal + Burst & Baseline resilience \\
TopologyZoo-2 & ISP Backbone & 230 & 410 & Cascading & Adversarial & Propagation analysis \\
CAIDA-AS & AS-level & 5,000+ & 20,000+ & Multi-node & Adversarial & Inter-domain stress \\
\bottomrule
\end{tabular}
}
\caption{We evaluate our framework across diverse network topologies summarized in Table 1, ranging from ISP-level backbone networks to AS-level inter-domain graphs.}
\label{tab:datasets}
\end{table*}

\textbf{Baselines.}
We compare against (1) rule-based fault injection with manually defined failures, (2) Digital Twin simulation without LLM-based scenario generation, and (3) historical failure replay without generative exploration.

\textbf{Evaluation Metrics.}
Evaluation focuses on standard network KPIs, including end-to-end latency, packet loss, reroute convergence time, congestion levels, and
the number of impacted nodes. Mitigation effectiveness is measured by relative improvement in these metrics after recovery actions.

\subsection{Main Results}

\begin{figure}[ht]
\centering
\includegraphics[width=0.37\textwidth]{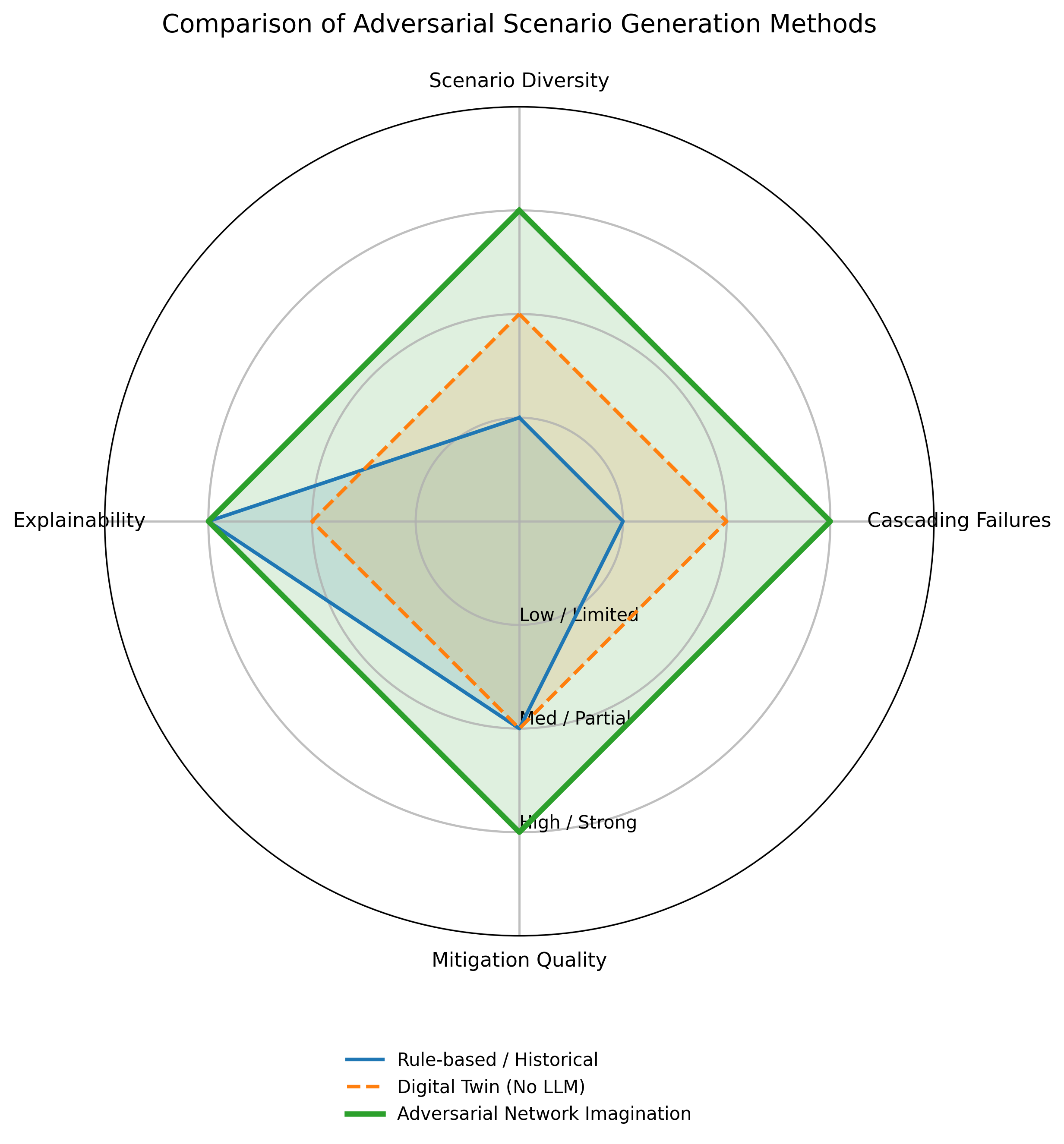} 
\caption{Qualitative comparison of adversarial scenario generation and mitigation effectiveness.}
\label{fig:MainResult}
\end{figure}

Figure~\ref{fig:MainResult} summarizes the comparative behavior of the proposed framework. Adversarial Network Imagination consistently
generates more diverse and complex failure scenarios than baseline approaches, including multi-component and cascading events that are not captured by rule-based or replay-based methods. Causal conditioning enables deeper propagation within the Digital Twin, resulting in more informative stress testing and stronger relative mitigation gains. These results demonstrate the benefit of closed-loop, causally grounded scenario generation for proactive resilience analysis.

\subsection{Ablation Study}

\begin{figure}[ht]
\centering
\includegraphics[width=0.45\textwidth]{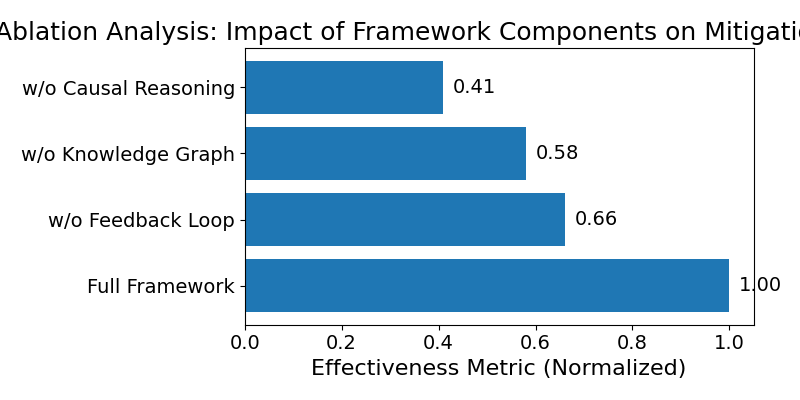}
\caption{Ablation study showing normalized performance impact of removing individual framework components. Scores are normalized relative to the full framework.}
\label{fig:ablationscore}
\end{figure}

To analyze the contribution of individual components, we conduct an ablation study summarized in Figure~\ref{fig:ablationscore}. Removing the Knowledge Graph substantially reduces scenario realism and propagation depth, while disabling causal conditioning weakens cascading behavior and mitigation effectiveness. Excluding simulation feedback leads to less adaptive refinement across iterations. Overall, the ablation results confirm that causal structure, dependency awareness, and closed-loop verification each play a critical role in generating high-impact adversarial failure scenarios.

\begin{figure}[ht]
\centering
\includegraphics[width=0.45\textwidth]{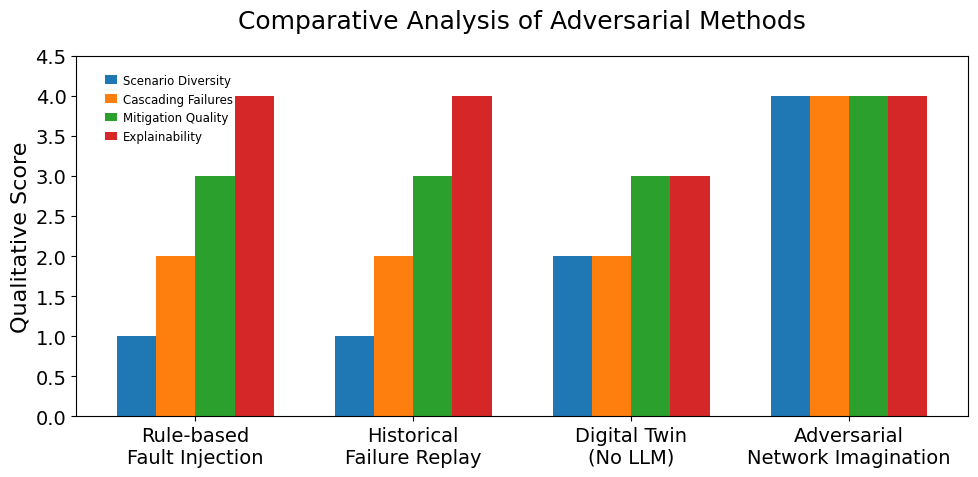}
\caption{Ablation Study Effectiveness.}
\label{fig:ablation_chart}
\end{figure}

\FloatBarrier

\subsection{Cross-Topology Generalization}

We further examine whether the framework generalizes across different network structures by applying it to multiple ISP topologies from the Internet Topology Zoo with varying sizes and connectivity patterns. The Causal LLM adapts scenario generation to topology-specific dependencies, producing distinct failure patterns for each network. This experiment demonstrates that the framework is not overfitted to a single topology and can support resilience analysis across heterogeneous telecom environments.

\subsection{Scenario Triggering and Digital Twin Execution}

The Causal LLM generates structured failure descriptions, e.g., ``Fiber link between B and C goes down,'' which are converted into machine-readable events (e.g., \texttt{"event\_type": "fiber\_link\_failure", "target": "B-C", "timestamp": 245.8}) and executed in the Digital Twin. The Digital Twin models routing, resource utilization, congestion, and service-level performance, applying stress conditions such as traffic surges, overloaded routers, and packet drops. Cascading effects are tracked, and metrics including latency, packet loss, reroute time, congestion, and impacted nodes—are recorded before and after mitigation to assess recovery effectiveness.

\subsection{Evaluation Scenarios}

We evaluate the framework under representative and adversarial failure conditions:

\begin{itemize}
    \item \textbf{Fiber link failures:} Evaluate routing convergence and congestion.
    \item \textbf{Router overloads:} Assess traffic redistribution and control-plane responsiveness.
    \item \textbf{Multi-node cascading outages:} Examine propagation depth of sequential dependent failures.
    \item \textbf{DDoS-style traffic spikes:} Test resilience against high-intensity traffic.
\end{itemize}

\begin{table}[h!]
\centering
\resizebox{\linewidth}{!}{%
\begin{tabular}{|l|l|l|}
\hline
\textbf{Scenario} & \textbf{Description} & \textbf{Measured KPIs} \\
\hline
Fiber link failure & Routing adjustments and congestion & Latency, packet loss, impacted nodes \\
Router overload & Traffic redistribution under stress & Latency, congestion, reroute time \\
Cascading outages & Sequential failures of dependent nodes & Latency, packet loss, impacted nodes \\
DDoS-style spike & High-intensity adversarial traffic & Latency, packet loss, congestion levels \\
\hline
\end{tabular}%
}
\caption{Evaluation scenarios and associated KPIs.}
\label{tab:evaluation_scenarios}
\end{table}

\section{Safety and Ethical Considerations}

Digital twin deployments in telecom networks raise important security and resilience considerations \citep{Salman2022}. To mitigate operational risk, all failure scenarios are generated and executed exclusively within a controlled simulation environment; no actions are applied to live production networks. Knowledge graph representations and LLM inputs are fully anonymized to protect sensitive infrastructure and customer information \citep{Gai2018}. While the proposed framework enables systematic exploration of adversarial failure scenarios, all generated outputs are intended to support analysis and planning rather than automated decision-making, and should be interpreted under human oversight.

\section{Conclusion}
This work presents an autonomous framework, \emph{Adversarial Network Imagination}, that integrates causal reasoning, generative LLMs, and Digital Twin simulation for proactive telecom mitigation. By systematically generating, simulating, and evaluating complex failure scenarios, the framework moves network management from a reactive to an anticipatory paradigm. Evaluation through diverse adversarial scenarios demonstrates the system’s ability to uncover vulnerabilities, measure cascading effects, and assess mitigation strategies effectively. Future developments will focus on real-data integration, reinforcement learning for automatic mitigation, and scaling to heterogeneous network environments, supporting resilient and reliable telecommunications infrastructure.

\section*{Limitations}
The proposed framework has several practical limitations. The fidelity of the Digital Twin depends on the availability and accuracy of network topology and resource data, and incomplete or outdated information can reduce simulation realism. Scalability remains a challenge, as large cascading failures and high-traffic stress tests increase computational cost. The Causal LLM may occasionally generate incomplete or imprecise scenarios that require validation before use. In addition, limited access to real operational datasets can constrain realism and generalization. Addressing these challenges will require continued refinement of simulation efficiency, scenario prioritization, and model robustness.

\bibliography{custom}

\appendix
\label{sec:appendix}

\appendix
\section{Appendix: Scenario Triggering and Event Representation}
\label{sec:appendix_triggering}

This appendix provides additional details on how adversarial failure scenarios are triggered and executed within the Digital Twin simulation. These details are included for clarity and reproducibility and are not required for understanding the main contributions of the paper.

\subsection{Failure Scenario Generation}

The Causal LLM generates structured natural-language descriptions of network failures grounded in dependency constraints from the Knowledge Graph. Each description specifies the failure type, affected components, and optional temporal or stress-related attributes. An example generated description is:

\begin{quote}
\emph{``A fiber link between routers B and C fails during peak traffic, causing rerouting and congestion on adjacent paths.''}
\end{quote}

This representation is designed to be explainable to human operators while remaining convertible to a machine-readable format.

\subsection{Event Encoding}

Generated scenarios are translated into structured events that can be executed by the Digital Twin. Events follow a simple schema that captures the failure type, target component, and activation time. An example event encoding is shown below:

\begin{verbatim}
{
  "event_type": "fiber_link_failure",
  "target": "B-C",
  "timestamp": 245.8,
  "severity": "high"
}
\end{verbatim}

This abstraction allows different failure types (e.g., link failures, router overloads, traffic spikes) to be handled uniformly within the simulation engine.

\subsection{Digital Twin Execution}

Once triggered, events are applied within the Digital Twin environment, which models routing behavior, resource utilization, congestion, and service-level performance. Stress conditions such as traffic surges, packet loss, or CPU exhaustion may be injected alongside the primary failure to simulate adversarial conditions.

The Digital Twin tracks both direct and cascading effects, recording metrics including latency, packet loss, reroute time, congestion levels, and the set of impacted nodes.

\subsection{Mitigation Feedback}

After simulation, candidate mitigation actions (ex, traffic rerouting or resource reallocation) are applied, and post-mitigation metrics are collected. These outcomes are fed back to the scenario generator, enabling iterative refinement of future failure scenarios.

All the triggering and execution occur exclusively within the simulated Digital Twin environment; no actions are applied to live production networks.

\subsection{Checklist and Reproducibility Clarifications}

\paragraph{Computational Experiments and Hyperparameters.}
The proposed framework does not involve training or fine-tuning of machine learning models. The large language model is used as a fixed component for structured scenario generation under causal constraints.
As such, no hyperparameter search or optimization is performed.

\paragraph{Artifacts and Documentation.}
The paper documents the structure of generated failure scenarios, event encodings, and their execution within the Digital Twin to support conceptual reproducibility, even though no executable artifacts are released.

\paragraph{Experimental Setup and Statistics.}
Evaluation is conducted through deterministic Digital Twin simulations of generated failure scenarios. Reported outcomes reflect system-level metrics (example., latency, congestion, packet loss) for each scenario, rather than averages over multiple stochastic training runs. Therefore, descriptive statistics such as variance or confidence intervals are not applicable.

\paragraph{Use of AI Assistants.}
A large language model is used exclusively to generate structured, human-interpretable descriptions of hypothetical failure scenarios consistent with the Knowledge Graph. The model is not used for network control, mitigation decision-making, or result evaluation. All simulations and measurements are executed within the Digital Twin environment.

\section{Discussion}
Integrating causal reasoning with generative LLMs and high-fidelity Digital Twin simulations enables proactive telecom resilience in ways that manual testing or rule-based fault injection cannot. Even minor network failures can cascade through complex dependencies, and adversarial scenario generation helps uncover latent vulnerabilities before they affect users. The closed-loop framework allows the LLM to iteratively refine scenario generation based on simulation feedback, improving its ability to produce operationally relevant, high-impact failure cases. While human supervision is still important for validation, the framework significantly reduces the need for constant manual oversight, supporting anticipatory rather than reactive network management.

\section{Future Work and Real World Impact}
Future work will focus on enhancing the realism, scalability, and automation of the framework. Integrating real operational data into Knowledge Graphs and Digital Twin models will improve fidelity and causal reasoning. Reinforcement learning techniques could automatically identify optimal mitigation strategies and accelerate scenario refinement. The framework can also be extended to multi-domain, cloud, and 5G networks, enabling cross-layer resilience analysis. Continuous feedback loops between simulation outcomes and generative models will allow networks to discover rare failure modes proactively and improve anticipatory response capabilities. The ultimate impact is a shift in operational practice: network operators can evaluate, anticipate, and mitigate risks before they manifest, reducing downtime and improving service reliability.

\end{document}